\documentclass[uplatex, a4paper,12pt]{article} 
\usepackage[utf8]{inputenc}
\usepackage{amsmath}
\usepackage{color}
\usepackage{amssymb, amsfonts, amsthm, amscd}
\usepackage{bm}
\usepackage{float}
\usepackage{caption}
\usepackage[subrefformat=parens]{subcaption}
\usepackage{hyperref}  % こちらを統一
\usepackage{graphicx}  % こちらを統一
\usepackage{bookmark}            % hyperrefの代替
\hypersetup{% hyperrefオプションリスト
  setpagesize=false,
  bookmarksnumbered=true,
  bookmarksopen=true,
  colorlinks=true,
  linkcolor=blue,
  citecolor=red,
}
\begin{document}

%%%%%%%%%
%%%%%%%%%
\thispagestyle{empty}
\vspace*{-15mm}

\begin{flushleft}
{\bf OUJ-FTC-12}\\

\end{flushleft}

%----------
%\baselineskip 1pt
%\begin{flushright}
%\begin{tabular}{l}
{\bf }\
%%%%%%%%%%%%%%%%%%%%%%%%%%

\vspace{15mm}

\begin{center}
{\Large\bf
Prediction of the Dark Fermion Mass using
Multicritical-Point Principle}

\baselineskip 18pt
\vspace{7mm}

%{\bf The OUJ Tokyo Bunkyo Field Theory Collaboration}

Yoshiki Matsuoka

\vspace{5mm}

{\it Nature and Environment, Faculty of Liberal Arts, The Open University of Japan, Chiba 261-8586, Japan\\

}

\end{center}

\vspace{3cm}

\begin{flushleft} 
Email: machia1805@gmail.com  
\end{flushleft}

%%%%%%%%%%%%%%%%%
\vspace{10mm}
%%%%%%%%%%%%%%%%%%%%%%%%%%%%%%%%%%%%%%%%
%%%%%                              %%%%%
%%%%%          Abstract            %%%%%
%%%%%                              %%%%%
%%%%%%%%%%%%%%%%%%%%%%%%%%%%%%%%%%%%%%%%
\begin{center}
\begin{minipage}{14cm}
\baselineskip 16pt
\noindent
%%%%%---------------------------------
\begin{abstract}
This paper proposes a method to determine the effective potential using the multicritical-point principle (MPP) under the additional scalar field.
The MPP is applied to the model in which a singlet dark fermion and a singlet real scalar field are added to the Standard Model (SM) to predict the dark fermion mass.
As a result, the dark fermion mass is predicted to be about $901$--$972$ GeV.

\end{abstract}

%%%%%----------------------------------
\end{minipage}
\end{center}

\vspace{0.5cm}
%%%%%%%%%%%
%%%%%%%%%%%%%%%
\newpage
\section{Introduction}
Although about a decade has passed since the discovery of the Higgs boson, many unsolved problems still exist\cite{higgs1,higgs2}. In particular, the nature of the dark matter (DM) \cite{dark} is still the major unsolved problem, and many people continue to challenge it using various approaches\cite{dark_re} such as theories\cite{TH1,TH2,TH3,TH4,TH5}, experiments\cite{EX1,EX2,EX3}, numerical computations\cite{NC}, and machine learnings\cite{ML1,ML2}.

About 20 years before the discovery of the Higgs, Froggatt and Nielsen proposed what they called the multicritical-point principle (MPP) and predicted the Higgs boson mass as $135\pm9$ GeV with $173\pm5$ GeV for the top quark mass in \cite{MPPoriginal,MPPoriginal2,MPPoriginal3}. So far, various proposals of the MPP have been applied\cite{kk1,kk2,kk3,mppeg1,mppeg2,mppeg3,mppeg4,kawai3,kawai1,kawai2}. The MPP is based on the idea that ``some of extreme values of the theory's effective potential should be degenerate.'' For $h$ taken as the Higgs field, the MPP means that the energy density tends to degenerate in two different vacuum states, $V_{\mathrm{\mathrm{eff}}}(h_{\mathrm{EW}})=V_{\mathrm{eff}}(M_{\mathrm{pl}})\simeq0,\
\frac{dV_{\mathrm{eff}}(h_{\mathrm{EW}})}{dh}=\frac{dV_{\mathrm{eff}}(M_{\mathrm{pl}})}{dh}\simeq0$, where $V_{\mathrm{\mathrm{eff}}}$ is the effective potential, $h_{\mathrm{EW}} = 246$ GeV and $M_{\mathrm{pl}}\simeq 10^{19}$ GeV. 
The first equation represents the equality of the energies of the two phases, which is the prerequisite for the coexistence of the two phases. It is important to note that the two values of the effective potential are given on so different scales.
In order to match the values of the effective potential at $h_{\mathrm{EW}}=246\mathrm{GeV}$ and $M_{\mathrm{pl}}\simeq 10^{19}\mathrm{GeV}$, which are clearly on significantly different scales, $V_{\mathrm{eff}}(M_{\mathrm{ pl}})$ must be about zero.

In this paper, we considered the MPP to be valid even for BSM(Beyond the Standard Model). So, we have attempted to predict the DM mass discussed in \cite{dark,dark_re} using MPP in the additinal scalar field as a simple extension of the SM, which is relatively easy to realize the MPP. We have applied the MPP to the effective potential in the two scalar fields including the Higgs field assuming the MPP to predict the DM mass. The result shows that the DM mass is WIMP(Weakly Interacting Massive Particle)-like.

We organize the paper as follows. In the next section, we review the MPP. In Section 3, we define our model and apply our MPP conditions. In Section 4, we compare our predictions with observations. In Section 5, the summary and the discussion are given. In Appendix A, we list the one-loop renormalization group equations (RGEs). In Appendix B, we calculate the annihilation cross section.

%%%%%%%%%%%%%%%%%%%%%%%%%
%%%%%%%%%%%%%%%%%%%%%%%%%
\section{Brief review of the MPP}
We briefly explain the MPP.
The MPP introduced in \cite{MPPoriginal} is to impose the following conditions on the SM:
\begin{eqnarray}\label{mppori}
V_{\mathrm{\mathrm{eff}}}(h_{\mathrm{EW}})=V_{\mathrm{eff}}(M_{\mathrm{pl}})\simeq0,\\
\label{mppori2}
\frac{dV_{\mathrm{eff}}(h_{\mathrm{EW}})}{dh}=\frac{dV_{\mathrm{eff}}(M_{\mathrm{pl}})}{dh}\simeq0
\end{eqnarray}
where $V_{\mathrm{\mathrm{eff}}}$ is the one-loop effective potential, $h_{\mathrm{EW}} = 246$ GeV and $M_{\mathrm{pl}}\simeq 10^{19}$ GeV. The reason why (\ref{mppori}) becomes approximately zero is that $h_{\mathrm{EW}}$ is so small compared to $M_\mathrm{pl}$.

We apply the dimensional regularization to $V_{\mathrm{\mathrm{eff}}}$ and shift the spacetime dimension from $4$ to $D$ and eliminate its poles in a way that depends on the energy scale $\mu$ and perform Wick rotation.
And we set the renormalization conditions as follows:
\begin{eqnarray}
\left.V_{\mathrm{eff}}\right|_{h=0} &=& 0,\\
\left.\frac{d^2V_{\mathrm{eff}}}{dh^2}\right|_{h=0} &=& 0,\\
\left.\frac{d^4V_{\mathrm{eff}}}{dh^4}\right|_{h=M_t} &=& 6\lambda(M_t),\\
\left.\frac{dV_{\mathrm{eff}}}{dh}\right|_{h=0}=\left.\frac{d^3V_{\mathrm{eff}}}{dh^3}\right|_{h=0} &=& 0
\end{eqnarray}
where $\lambda$ is the Higgs quartic coupling constant and $M_t$ is the top quark mass.
Then we use $\mathrm{\overline{MS}}$ scheme \cite{higgs3} to remove the $\lim_{D\to4}(1/(4-D))$ poles, the Euler's constant $\gamma_E$, and $\ln(4\pi)$ from the one-loop effective potential $V_{\mathrm{eff}}(h)$. In other words, we simply subtract these three values from the one-loop effective potential.

In the SM, the one-loop effective potential in Landau gauge\footnote{Thus, the mass of the gauge fields takes on the overall factor of $\lim_{D\to4}(3-(4-D))$.} is given by
\begin{eqnarray}
V_{\mathrm{eff}}\left(h(\mu), \mu(t)\right) &=& \frac{\lambda(\mu)}{4}h^4(\mu)
+\frac{1}{64\pi^2}\left(3\lambda(\mu) h^2(\mu)\right)^2\left(\ln\frac{3\lambda(\mu) h^2(\mu)}{\mu^2(t)}-\frac{3}{2}\right)\nonumber\\
&+&\frac{3}{64\pi^2}\left(\sqrt{\lambda(\mu)}h(\mu)\right)^4\left(\ln\frac{\lambda(\mu) h^2(\mu)}{\mu^2(t)}-\frac{3}{2}\right)\nonumber\\
&+&\frac{3\times2}{64\pi^2}\left(\frac{g_2(\mu)h(\mu)}{2}\right)^4\left(\ln\frac{\left(g_2(\mu)h(\mu)\right)^2}{4\mu^2(t)}-\frac{5}{6}\right)\nonumber\\
&+&\frac{3}{64\pi^2}\left(\frac{\sqrt{g_2^2(\mu)+g_Y^2(\mu)}}2h(\mu)\right)^4\left(\ln\frac{\left(g_2^2(\mu)+g_Y^2(\mu)\right)h^2(\mu)}{4\mu^2(t)}-\frac{5}{6}\right)\nonumber\\
&-&\frac{4\times3}{64\pi^2}\left(\frac{y_t(\mu)h(\mu)}{\sqrt{2}}\right)^4\left(\ln\frac{\left(y_t(\mu)h(\mu)\right)^2}{2\mu^2(t)}-\frac{3}{2}\right)
\end{eqnarray}
where $\mu$ is the renormalization scale, $\lambda(\mu),y_t(\mu), g_Y(\mu), g_2(\mu)$ are the Higgs quartic coupling constant, the top Yukawa coupling constant, and the $U(1)_Y$ and $SU(2)_L$ gauge coupling constants, each being  a function of $\mu(t)$, and
\begin{eqnarray}
\mu(t) \equiv M_te^{t}
\end{eqnarray}
where $M_t$ is the top quark mass and $t$ is a certain parameter. We consider only the contribution of the top quark, which has the large Yukawa coupling constant among the quarks when we consider the one-loop effective potential. $h(\mu)$ is the renormalized running field. The renormalized running field is given by
\begin{eqnarray}\label{RRF}
h(\mu) = he^{\Gamma(\mu)}
\end{eqnarray}
where $h$ is the bare field and $\Gamma(\mu)$ is the wave function renormalization. We do not distinguish between the bare field $h$ and the renormalized running field $h(\mu)$ because $\Gamma(\mu)$ is so small.
In \cite{MPPoriginal}, Froggatt and Nielsen simply choose $\mu(t)=h(\mu)$. And they predicted the Higgs boson mass as $135\pm9$ GeV using the MPP conditions (\ref{mppori}), (\ref{mppori2}) on the Planck scale.

%%%%%%%%%%%%%%
%%%%%%%%%%%%%%
\section{The Model}
We consider our model and how the MPP is applied.
Now we consider the following renormalizable Lagrangian such that the fermion number of the singlet dark fermion is conserved:
\begin{eqnarray}\label{model}
\mathcal{L}\equiv\mathcal{L}_{\mathrm{SM}}&+&\frac{1}{2}(\partial_\mu S)^2-\frac{\lambda_{\mathrm{DM}}}{4!}S^4+i\overline{\chi}\gamma^\mu \partial_{\mu}\chi-\frac{y_\chi}{\sqrt{2}}\overline{\chi}S\chi
+\frac{1}{2}\kappa(H^\dagger H)S^2
\end{eqnarray}
where $H,S,\chi$ are the Higgs field without the mass term, the singlet real scalar, and the singlet dark Dirac fermion\footnote{We considered terms in $H$ and $S$ with $+$(boson) and $-$(fermion) contributions with respect to the effective potential.}\cite{flux}.
$\lambda_{\mathrm{DM}}(\mu), \kappa(\mu), y_\chi(\mu)$ are the real scalar quatric coupling constant, the scalar interaction coupling constant, and the dark Yukawa coupling constant and they are a function of $\mu(t)$.
$H$ and $S$ can be written as follows:
\begin{eqnarray}
H &=& \frac{1}{\sqrt{2}}\begin{pmatrix}
   0 \\
   h(\mu)
\end{pmatrix}, \\ S &=& s(\mu).
\end{eqnarray}
where $h(\mu)$ and $s(\mu)$ are vacuum expectation values (VEVs).
We set the renormalization conditions as follows:
\begin{eqnarray}
\left.V_{\mathrm{eff}}\right|_{s=0,  h=0}&=& 0,\\
\left.\frac{\partial^2V_{\mathrm{eff}}}{\partial h^2}\right|_{s=0, h=0} &=& 0,\\
\left.\frac{\partial^4V_{\mathrm{eff}}}{\partial h^4}\right|_{s=0, h=M_t} &=& 6\lambda(M_t),\\
\left.\frac{\partial^2V_{\mathrm{eff}}}{\partial s^2}\right|_{s=0, h=0} &=& 0,\\
\left.\frac{\partial^4V_{\mathrm{eff}}}{\partial s^4}\right|_{s=M_t, h=0}  &=& \lambda_{\mathrm{DM}}(M_t),\\
\left.\frac{\partial^4V_{\mathrm{eff}}}{\partial s^2\partial h^2}\right|_{s=M_t, h=M_t}  &=& -\kappa(M_t),\\
\left.\frac{\partial V_{\mathrm{eff}}}{\partial h}\right|_{s=0, h=0}=\left.\frac{\partial V_{\mathrm{eff}}}{\partial s}\right|_{s=0, h=0}&=&0,\\
\left.\frac{\partial^3 V_{\mathrm{eff}}}{\partial h^3}\right|_{s=0, h=0}=\left.\frac{\partial^3 V_{\mathrm{eff}}}{\partial s^3}\right|_{s=0, h=0}=\left.\frac{\partial^3 V_{\mathrm{eff}}}{\partial h^2\partial s}\right|_{s=0, h=0}=\left.\frac{\partial^3 V_{\mathrm{eff}}}{\partial h\partial s^2}\right|_{s=0, h=0}&=& 0,\\
\left.\frac{\partial^2V_{\mathrm{eff}}}{\partial s\partial h}\right|_{s=0, h=0}=
\left.\frac{\partial^4V_{\mathrm{eff}}}{\partial s^3\partial h}\right|_{s=0, h=0}=
\left.\frac{\partial^4V_{\mathrm{eff}}}{\partial s\partial h^3}\right|_{s=0, h=0}&=&0.
\end{eqnarray}
The one-loop effective potential is as follows:
\begin{eqnarray}
V_{\mathrm{eff}}\left(s(\mu),h(\mu),\mu(t) \right) &=& \frac{\lambda(\mu)}{4}h^4(\mu)+\frac{\lambda_{\mathrm{DM}}(\mu)}{4!}s^4(\mu)-\frac{1}{4}\kappa(\mu) h^2(\mu)s^2(\mu)\nonumber\\
&+&\frac{1}{64\pi^2}\left(\left(3\lambda(\mu) h^2(\mu)-\frac{\kappa(\mu)}{2}s^2(\mu)\right)^2+\bigg(\kappa(\mu)h(\mu)s(\mu)\bigg)^2\right)\nonumber\\
&\times&\left(\ln\frac{3\lambda(\mu) h^2(\mu)-\frac{\kappa(\mu)}{2}s^2(\mu)}{\mu^2(t)}-\frac{3}{2}\right)\nonumber\\
&+&\frac{3}{64\pi^2}\left(\lambda(\mu) h^2(\mu)-\frac{\kappa(\mu)}{2}s^2(\mu)\right)^2\left(\ln\frac{\lambda(\mu) h^2(\mu)-\frac{\kappa(\mu)}{2}s^2(\mu)}{\mu^2(t)}-\frac{3}{2}\right)\nonumber\\
&+&\frac{3\times2}{64\pi^2}\left(\frac{g_2(\mu)h(\mu)}{2}\right)^4\left(\ln\frac{\left(g_2(\mu)h(\mu)\right)^2}{4\mu^2(t)}-\frac{5}{6}\right)\nonumber\\
&+&\frac{3}{64\pi^2}\left(\frac{\sqrt{g_2^2(\mu)+g_Y^2(\mu)}}2h(\mu)\right)^4\left(\ln\frac{\left(g_2^2(\mu)+g_Y^2(\mu)\right)h^2(\mu)}{4\mu^2(t)}-\frac{5}{6}\right)\nonumber\\
&-&\frac{4\times3}{64\pi^2}\left(\frac{y_t(\mu)h(\mu)}{\sqrt{2}}\right)^4\left(\ln\frac{\left(y_t(\mu)h(\mu)\right)^2}{2\mu^2(t)}-\frac{3}{2}\right)\nonumber\\
&+&\frac{1}{64\pi^2}\left(\left(\frac{\lambda_{\mathrm{DM}}(\mu)}{2}s^2(\mu)-\frac{\kappa(\mu)}{{2}}h^2(\mu)\right)^2+\bigg(\kappa(\mu)h(\mu)s(\mu)\bigg)^2\right)\nonumber\\
&\times&\left(\ln\frac{\frac{\lambda_{\mathrm{DM}}(\mu)}{2}s^2(\mu)-\frac{\kappa(\mu)}{{2}}h^2(\mu)}{\mu^2(t)}-\frac{3}{2}\right)\nonumber\\
&-&\frac{4}{64\pi^2}\left(\frac{y_\chi(\mu)s(\mu)}{\sqrt{2}}\right)^4\left(\ln\frac{\left(y_\chi(\mu)s(\mu)\right)^2}{2\mu^2(t)}-\frac{3}{2}\right)\nonumber\\
&-&\frac{1}{64\pi^2}\left(\kappa(\mu)h(\mu)s(\mu)\right)\nonumber\\
&\times&\left(\left(3\lambda(\mu) h^2(\mu)-\frac{\kappa(\mu)}{2}s^2(\mu)\right)+\left(\frac{\lambda_{\mathrm{DM}}(\mu)}{2}s^2(\mu)-\frac{\kappa(\mu)}{{2}}h^2(\mu)\right)\right)\nonumber\\
&\times&\left(\ln\frac{\kappa(\mu)h(\mu)s(\mu)}{\mu^2(t)}-\frac{3}{2}\right)
\end{eqnarray}
where $s(\mu)$ and $h(\mu)$ are the renormalized running fields (\ref{RRF}).

In this case, the singlet real scalar field $s(\mu)$ and the Higgs field $h(\mu)$ are mixing at tree-level.
The mixing angle $\theta$ is defined as 
\begin{eqnarray}
\left(\begin{array}{c}
h_1(\mu) \\
h_2(\mu)
\end{array}\right)=\left(\begin{array}{cc}
\cos \theta & -\sin \theta \\
\sin \theta & \cos \theta
\end{array}\right)\left(\begin{array}{l}
s(\mu) \\
h(\mu)
\end{array}\right).
\end{eqnarray}
where the off-diagonal terms appear due to the interaction term by $\kappa(\mu)$. Diagonalized forms can be to eliminate the off-diagonal terms in the anomalous dimensions. $\theta$ is given by
\begin{eqnarray}
\tan(2\theta) = \frac{2\kappa h_{\mathrm{EW}}s_{\mathrm{DM}}}{M_{h_1}^2-M_{h_2}^2}
\end{eqnarray}
where $h_{\mathrm{EW}}$ and $s_{\mathrm{DM}}$ are VEVs on the Electroweak scale. $M_{h_1}$ and $M_{h_2}$ are the singlet real scalar mass and the SM-like Higgs mass.  These are given by
\begin{eqnarray}
M_{h_{1,2}}^2=\frac{1}{2}(M_{s}^2+M_{h}^2)\mp\frac{1}2\sqrt{(M_s^2-M_h^2)^2+4\kappa^2h_{\mathrm{EW}}^2s_{\mathrm{DM}}^2}
\end{eqnarray}
where $M_{s}=\sqrt{\frac{\lambda_{\mathrm{DM}}(M_t)}{2!}}s_{\mathrm{DM}}$ and $M_{h}=\sqrt{2\lambda(M_t)}h_{\mathrm{EW}}$ and we consider $\theta\ll0.1$.

The one-loop effective potential $V_{\mathrm{eff}}\left(h_1(\mu),h_2(\mu),\mu(t) \right)$ is as follows:
\begin{eqnarray}
V_{\mathrm{eff}}\left(h_1(\mu),h_2(\mu),\mu(t) \right) &=& \frac{\lambda(\mu)}{4}(-h_1(\mu)\sin\theta+h_2(\mu)\cos\theta)^4+\frac{\lambda_{\mathrm{DM}}(\mu)}{4!}(h_1(\mu)\cos\theta+h_2(\mu)\sin\theta)^4\nonumber\\
&-&\frac{1}{4}\kappa(\mu) (-h_1(\mu)\sin\theta+h_2(\mu)\cos\theta)^2(h_1(\mu)\cos\theta+h_2(\mu)\sin\theta)^2+\frac{1}{64\pi^2}\nonumber\\
&\times&\Biggl(\Bigg(3\lambda(\mu) (-h_1(\mu)\sin\theta+h_2(\mu)\cos\theta)^2
-\frac{\kappa(\mu)}{2}(h_1(\mu)\cos\theta+h_2(\mu)\sin\theta)^2\Bigg)^2\nonumber\\
&+&\Bigg(\kappa(\mu)\left(-h_1(\mu)\sin\theta+h_2(\mu)\cos\theta\right)\left(h_1(\mu)\cos\theta+h_2(\mu)\sin\theta\right)\Bigg)^2\Biggl)\nonumber\\
&\times&\Bigg(\ln\frac{3\lambda(\mu) (-h_1(\mu)\sin\theta+h_2(\mu)\cos\theta)^2-\frac{\kappa(\mu)}{2}(h_1(\mu)\cos\theta+h_2(\mu)\sin\theta)^2}{\mu^2(t)}\nonumber\\
&-&\frac{3}{2}\Bigg)+3\Bigg(\lambda(\mu) (-h_1(\mu)\sin\theta+h_2(\mu)\cos\theta)^2\nonumber\\
&-&\frac{\kappa(\mu)}{2}(h_1(\mu)\cos\theta+h_2(\mu)\sin\theta)^2\Bigg)^2\nonumber\\
&\times&\Bigg(\ln\frac{\lambda(\mu) (-h_1(\mu)\sin\theta+h_2(\mu)\cos\theta)^2-\frac{\kappa(\mu)}{2}(h_1(\mu)\cos\theta+h_2(\mu)\sin\theta)^2}{\mu^2(t)}\nonumber\\
&-&\frac{3}{2}\Bigg)+\frac{3\times2}{64\pi^2}\left(\frac{g_2(\mu)(-h_1(\mu)\sin\theta+h_2(\mu)\cos\theta)}{2}\right)^4\nonumber\\
&\times&\left(\ln\frac{\left(g_2(\mu)(-h_1(\mu)\sin\theta+h_2(\mu)\cos\theta)\right)^2}{4\mu^2(t)}-\frac{5}{6}\right)\nonumber\\
&+&\frac{3}{64\pi^2}\left(\frac{\sqrt{g_2^2(\mu)+g_Y^2(\mu)}}2(-h_1(\mu)\sin\theta+h_2(\mu)\cos\theta)\right)^4\nonumber\\
&\times&\left(\ln\frac{\left(g_2^2(\mu)+g_Y^2(\mu)\right)(-h_1(\mu)\sin\theta+h_2(\mu)\cos\theta)^2}{4\mu^2(t)}-\frac{5}{6}\right)\nonumber\\
&-&\frac{4\times3}{64\pi^2}\left(\frac{y_t(\mu)(-h_1(\mu)\sin\theta+h_2(\mu)\cos\theta)}{\sqrt{2}}\right)^4\nonumber\\
&\times&\left(\ln\frac{\left(y_t(\mu)(-h_1(\mu)\sin\theta+h_2(\mu)\cos\theta)\right)^2}{2\mu^2(t)}-\frac{3}{2}\right)\nonumber
\end{eqnarray}
\begin{eqnarray}
&+&\frac{1}{64\pi^2}\Biggl(\Bigg(\frac{\lambda_{\mathrm{DM}}(\mu)}{2}(h_1(\mu)\cos\theta+h_2(\mu)\sin\theta)^2-\frac{\kappa(\mu)}{2}(-h_1(\mu)\sin\theta+h_2(\mu)\cos\theta)^2\Bigg)^2\nonumber\\
&+&\Bigg(\kappa(\mu)\left(-h_1(\mu)\sin\theta+h_2(\mu)\cos\theta\right)\left(h_1(\mu)\cos\theta+h_2(\mu)\sin\theta\right)\Bigg)^2\Biggl)\nonumber\\
&\times&\Bigg(\ln\frac{\frac{\lambda_{\mathrm{DM}}(\mu)}{2}(h_1(\mu)\cos\theta+h_2(\mu)\sin\theta)^2-\frac{\kappa(\mu)}{{2}}(-h_1(\mu)\sin\theta+h_2(\mu)\cos\theta)^2}{\mu^2(t)}\nonumber\\
&-&\frac{3}{2}\Bigg)-\frac{4}{64\pi^2}\left(\frac{y_\chi(\mu)(h_1(\mu)\cos\theta+h_2(\mu)\sin\theta)}{\sqrt{2}}\right)^4\nonumber\\
&\times&\left(\ln\frac{\left(y_\chi(\mu)(h_1(\mu)\cos\theta+h_2(\mu)\sin\theta)\right)^2}{2\mu^2(t)}-\frac{3}{2}\right)\nonumber\\
&+&\frac{-1}{64\pi^2}\Bigg(\kappa(\mu)\left(-h_1(\mu)\sin\theta+h_2(\mu)\cos\theta\right)\left(h_1(\mu)\cos\theta+h_2(\mu)\sin\theta\right)\Bigg)\nonumber\\
&\times&\Biggl(\Bigg(3\lambda(\mu) (-h_1(\mu)\sin\theta+h_2(\mu)\cos\theta)^2
-\frac{\kappa(\mu)}{2}(h_1(\mu)\cos\theta+h_2(\mu)\sin\theta)^2\Bigg)\nonumber\\
&+&\Bigg(\frac{\lambda_{\mathrm{DM}}(\mu)}{2}(h_1(\mu)\cos\theta+h_2(\mu)\sin\theta)^2-\frac{\kappa(\mu)}{2}(-h_1(\mu)\sin\theta+h_2(\mu)\cos\theta)^2\Bigg)\Biggl)\nonumber\\
&\times&\Bigg(\ln\frac{\kappa(\mu)(-h_1(\mu)\sin\theta+h_2(\mu)\cos\theta)(h_1(\mu)\cos\theta+h_2(\mu)\sin\theta)}{\mu^2(t)}-\frac{3}{2}\Bigg).
\end{eqnarray}
We consider this one-loop effective potential.

First, in order to obtain $s_{\mathrm{DM}}$, we consider the following VEV conditions on the Electroweak scale\cite{CW} around $\mu = M_t$\footnote{We consider something similar to the Coleman-Weinberg mechanism.}:
\begin{eqnarray}
\left.\frac{\partial V_{\mathrm{eff}}}{\partial h_1}\right|_{h_1=s_{\mathrm{DM}}, h_2=h_{\mathrm{EW}}}&=&\left.\frac{\partial V_{\mathrm{eff}}}{\partial h_2}\right|_{h_1=s_{\mathrm{DM}}, h_2=h_{\mathrm{EW}}}=0.
\end{eqnarray}

Next, we consider the MPP conditions on the Planck scale around $\mu\simeq M_{\mathrm{pl}}$ in order to get the three parameters $\lambda_{\mathrm{DM}}(\mu), y_\chi(\mu), \kappa(\mu)$ as follows:
\begin{eqnarray}\label{conditions}
\left.V_{\mathrm{eff}}\right|_{h_1=s_{\mathrm{pl}}, h_2=h_{\mathrm{pl}}}&=&0,\\
\label{conditions2}
\left.\frac{dV_{\mathrm{eff}}}{d\mu}\right|_{h_1=s_{\mathrm{pl}}, h_2=h_{\mathrm{pl}}}=\left.\frac{\partial V_{\mathrm{eff}}}{\partial h_1}\frac{dh_1}{d\mu}\right|_{h_1=s_{\mathrm{pl}}, h_2=h_{\mathrm{pl}}}+\left.\frac{\partial V_{\mathrm{eff}}}{\partial h_2}\frac{dh_2}{d\mu}\right|_{h_1=s_{\mathrm{pl}}, h_2=h_{\mathrm{pl}}}+\cdots&=&0.
\end{eqnarray}
We search for parameters $\lambda_{\mathrm{DM}}(\mu), y_\chi(\mu), \kappa(\mu)$ that satisfy these conditions. For this purpose, we consider the one-loop RGEs [\ref{firstappendix}]. Also, we find $s_{\mathrm{pl}}$ and $h_{\mathrm{pl}}$ such that the effective potential is minimized around $\mu\simeq M_{\mathrm{pl}}$.

We refer to \cite{cern,PDG}.
For the top quark mass $M_t=172.69\pm0.30\ \mathrm{GeV},$ the strong gauge coupling constant $\alpha(M_Z)=0.1179\pm0.0010$\cite{PDG}, we numerically found the parameters which approximately satisfy the VEV and the MPP conditions as follows:
\begin{eqnarray}\label{mpp}
 2550\ \mathrm{GeV}\lesssim &s_{\mathrm{DM}}&\lesssim2750\ \mathrm{GeV},\\
0.392\lesssim\lambda_{\mathrm{DM}}(M_t)\lesssim0.395,\ &y_\chi(M_t)&\simeq0.5,\ \kappa(M_t)\simeq0.008
\end{eqnarray}
where $\frac{M_{\mathrm{pl}}}{0.276}\lesssim s_{\mathrm{pl}} \lesssim \frac{M_{\mathrm{pl}}}{0.255}$ and $h_{\mathrm{pl}} \simeq \frac{M_{\mathrm{pl}}}{0.23}$ around $\mu\simeq M_{\mathrm{pl}}$ and we used $M_t$ and $\alpha(M_Z)$ to determine the values of the coupling constants in the SM. $\lambda_{\mathrm{DM}}(M_t)$, $s_{\mathrm{DM}}$, and $s_{\mathrm{pl}}$ inequality signs come from the standard deviations of $M_t$ and $\alpha(M_Z)$ and we consider $\theta\ll0.1$.

\begin{figure}[H]
 \begin{center}
 \includegraphics[width=100mm]{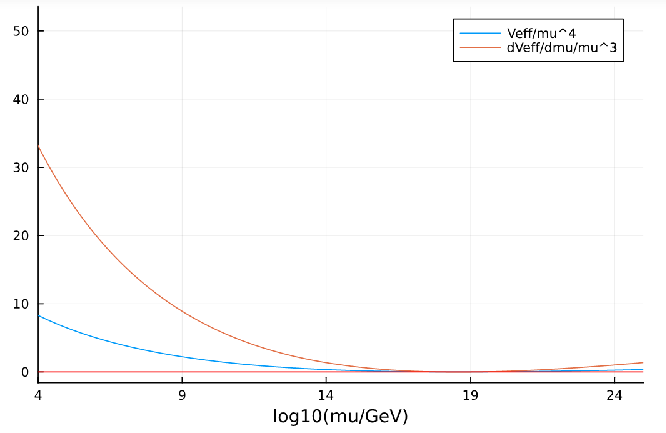}
 \caption{The $x$-axis and $y$-axis represent $\log_{10}(\frac{\mu(t)}{\mathrm{GeV}})$ and the value of each line. The blue line and orange line are $\frac{V_{\mathrm{eff}}}{\mu^4}$ and $\frac{1}{\mu^3}\frac{dV_{\mathrm{eff}}}{d\mu}$ in the case of $M_t=172.69\ \mathrm{GeV},\ \alpha(M_Z)=0.1189, \ \lambda_{\mathrm{DM}}(M_t) = 0.394$,\ $h_1(\mu)=\frac{\mu}{0.27}$, and $h_2(\mu) = \frac{\mu}{0.23}$. The red line represents zero on the $y$-axis.}\label{fig1}
 \end{center}
\end{figure}

Next, we check the MPP conditions on the Planck scale. We use $\frac{\mu}{0.276}\lesssim h_1(\mu) \lesssim \frac{\mu}{0.255}$ and $h_2(\mu) \simeq \frac{\mu}{0.23}$ since we are interested in the Planck scale. In the case of $M_t=172.69\ \mathrm{GeV},\ \alpha(M_Z)=0.1189, \ \lambda_{\mathrm{DM}}(M_t) = 0.394$,\ $h_1(\mu)=\frac{\mu}{0.27}$, and $h_2(\mu)=\frac{\mu}{0.23}$, the numerical results of (\ref{conditions}) and (\ref{conditions2}) are shown in Figure \ref{fig1}. 
We can see that these values approximately satisfy the MPP conditions.

\begin{figure}[H]
 \begin{center}
 \includegraphics[width=100mm]{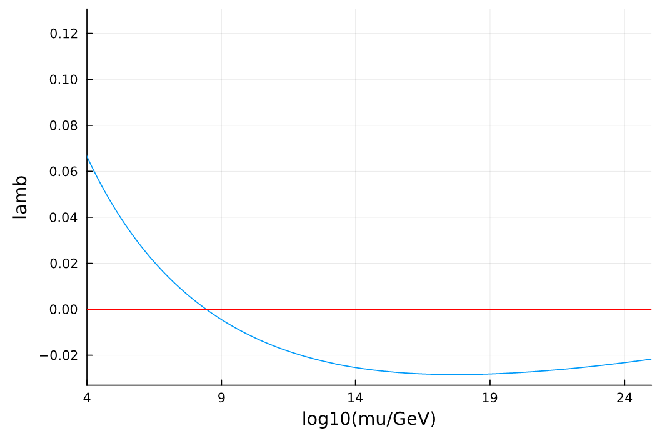}
 \caption{The $x$-axis and $y$-axis represent $\log_{10}(\frac{\mu(t)}{\mathrm{GeV}})$ and the value of $\lambda(\mu)$ in the case of $M_t=172.69\ \mathrm{GeV},\ \alpha(M_Z)=0.1189, \ \lambda_{\mathrm{DM}}(M_t) = 0.394$,\ $h_1(\mu)=\frac{\mu}{0.27}$, and $h_2(\mu)=\frac{\mu}{0.23}$. The red line represents zero on the $y$-axis.}\label{fig2}
 \end{center}
\end{figure}

Figure \ref{fig2} shows $\lambda(\mu)$ for the same parameters as Figure \ref{fig1} in the case of $M_t=172.69\ \mathrm{GeV}$,  $\alpha(M_Z)=0.1189, \ \lambda_{\mathrm{DM}}(M_t) = 0.394$,\ $h_1(\mu) = \frac{\mu}{0.27}$, and $h_2(\mu) = \frac{\mu}{0.23}$.
In other words, we check the vacuum stability, so to speak, whether $\lambda(\mu)$ is negative up to the Planck scale. $\lambda(\mu)$ reaches zero before it directly reaches zero on the Planck scale. The vacuum stability remains meta-stable as in the SM.

%%%%%%%%%%%%%%%%%
%%%%%%%%%%%%%%%%%%%%%%%%%%%%%%%
\section{Calculations of each value}
In this section, we calculate the masses, the decay rate, and the annihilation cross sections.
The singlet real scalar mass is
\begin{eqnarray}
1129\ \mathrm{GeV}\lesssim M_{h_1}\lesssim1222\ \mathrm{GeV}.
\end{eqnarray}
The DM mass is
\begin{eqnarray}
901\ \mathrm{GeV}\lesssim M_{\chi}\lesssim972 \ \mathrm{GeV}
\end{eqnarray}
where $M_\chi = \frac{y_\chi(M_t)}{\sqrt{2}}(h_1\cos\theta+h_2\sin\theta)$.
In order for our scenario to explain the observed relic density of DM 
($\Omega h^2 \sim 0.12$) and satisfy the current constraints from direct 
detection experiments, it is necessary to satisfy the condition $\theta < 0.1$ in $\mathrm{GeV}$--$\mathrm{TeV}$ scales, 
except in the region where $M_{h_1} \simeq M_{h_2}$\cite{dark_re}. In this case, the 
spin-independent elastic scattering cross-section $\sigma_{\chi N}^{\mathrm{SI}}$ for direct detection is 
constrained to $\sigma_{\chi N}^{\mathrm{SI}} \lesssim 10^{-18} \, \mathrm{GeV}^{-2}$. The spin-independent elastic scattering cross section between the DM and the (xenon) nucleus for direct detection is given by
\begin{eqnarray} 
\sigma_{\chi N}^{\mathrm{SI}}\simeq \frac{y_{\chi}^2M_{\chi}^2M_N^2}{32\pi(M_{\chi}+M_N)^2A^2h_{\mathrm{EW}}^2}\cos^2\theta\sin^2\theta\left(\frac{1}{M_{h_1}^2}-\frac{1}{M_{h_2}^2}\right)^2\times\left(Zf_p+(A-Z)f_n\right)^2
\end{eqnarray}
where $Z, A-Z$ are the numbers of protons and neutrons in the detector (xenon) nucleus, $M_N \simeq A M_p=131M_p$ is nucleus mass, $M_p$ is proton mass, and $f_p\simeq f_n \simeq 0.28M_p$.
Our mass range satisfies this constraint. This is a WIMP-like DM.\footnote{This result effectively reproduces the approximate equation for the DM relic abundunce, $\Omega_{\mathrm{DM}}h^2\sim0.1({M_{\chi}^2\over O(1)\mathrm{TeV}^2})(\frac{O(1)^4}{g_2^4})$. The Features of WIMPs are that it is so stable, the interactions are sufficiently weak, it is cold, and there is a suitable creation mechanism in the early universe.}

The singlet real scalar boson decays to the SM-like Higgs bosons:
\begin{eqnarray}
\Gamma(h_1\rightarrow h_2h_2)\simeq\frac{\kappa^2s_{\mathrm{DM}}^2}{32\pi M_{h_1}}\left(1-\frac{4M_{h_2}^2}{M_{h_1}^2}\right)^{\frac{1}2}.
\end{eqnarray}
The LHC searches for Higgs-like scalars. At present, the mixing angle $\theta$ is restricted to $\theta\lesssim0.3$.
In our model with the application of the MPP conditions, the mixing angle $\theta$ is limited to $\theta\ll0.1$. This satisfies the observation limits.

Next, we calculate the annihilation cross section between the DM and each SM particle because the DM can pair annihilate into a pair of the SM particles.
If this DM can pair annihilate into a pair of the SM fermions,
then the annihilation cross section is
\begin{eqnarray}\label{CS_ex}
(\sigma v_{\mathrm{rel}})_{\chi\overline{\chi}\to f\overline{f}} &\simeq& \frac{y_\chi^2y_f^2}{32\pi M_\chi^2}(\sin2\theta)^2\left(M_\chi^2\left(M_\chi^2-M_f^2\right)\right)\nonumber\\
&\times&\left[\frac{1}{4M_\chi^2-M_{h_1}^2}-\frac{1}{4M_\chi^2-M_{h_2}^2}\right]^2\left(1-\frac{M_f^2}{M_\chi^2}\right)^\frac{1}2.
\end{eqnarray}
For the cases of pair annihilations into a pair of the SM massive gauge bosons, the annihilation cross sections are
\begin{eqnarray}
(\sigma v_{\mathrm{rel}})_{\chi\overline{\chi}\to W^+W^-}&\simeq& \frac{y_\chi^2g_2^2}{128\pi M_\chi^2}(\sin2\theta)^2\left(4M_\chi^4-4M_W^2M_\chi^2+3M_W^4\right)\nonumber \\
&\times&\left[\frac{1}{4M_\chi^2-M_{h_1}^2}-\frac{1}{4M_\chi^2-M_{h_2}^2}\right]^2\left(1-\frac{M_W^2}{M_\chi^2}\right)^\frac{1}2,\\
(\sigma v_{\mathrm{rel}})_{\chi\overline{\chi}\to ZZ}&\simeq& \frac{y_\chi^2\left(g_2^2+g_Y^2\right)}{256\pi M_\chi^2}(\sin2\theta)^2\left(4M_\chi^4-4M_Z^2M_\chi^2+3M_Z^4\right)\nonumber \\
&\times&\left[\frac{1}{4M_\chi^2-M_{h_1}^2}-\frac{1}{4M_\chi^2-M_{h_2}^2}\right]^2\left(1-\frac{M_Z^2}{M_\chi^2}\right)^\frac{1}2.
\end{eqnarray}
Also, for the cases of pair annihilations into a pair of the SM-like Higgs bosons, the annihilation cross sections are
\begin{eqnarray}
(\sigma v_{\mathrm{rel}})_{\chi\overline{\chi}\to h_2h_2}&\simeq& \frac{y_\chi^2\kappa^2 s_{\mathrm{DM}}^2\left(2M_\chi^2-M_{h_2}^2\right)^2\left(8M_{\chi}^2-2M_{h_2}^2\right)^2}{256\pi\left(8M_\chi^4-6M_\chi^2M_{h_2}^2+M_{h_2}^4\right)^2\left(4M_\chi^2-M_{h_1}^2\right)^2}\times\left(1-\frac{M_{h_{2}}^2}{M_\chi^2}\right)^\frac{1}2.
\end{eqnarray}
We calculated the pair annihilation into a pair of the SM particles for each case.
%%%%%%%%%%%%%%%%%%%%%%%
%%%%%%%%%%%%%%%%%%%%%%%

\section{Summary and Discussion}
We discussed the MPP using the additinal scalar field. We proposed the MPP based on the model where the singlet dark fermion and the singlet real scalar field are added to the SM.
We placed four conditions in the model and determined six parameters $s_{\mathrm{DM}},\ s_{\mathrm{pl}},\ h_{\mathrm{pl}},\  \lambda_{\mathrm{DM}},\ \kappa,\ y_\chi$.  

In this case, for $M_t=172.69\pm0.30\ \mathrm{GeV}$ and $\alpha(M_Z)=0.1179\pm0.0010$, we found $2550\ \mathrm{GeV}\lesssim s_{\mathrm{DM}}\lesssim2750\ \mathrm{GeV},\ \frac{M_{\mathrm{pl}}}{0.276}\lesssim s_{\mathrm{pl}} \lesssim \frac{M_{\mathrm{pl}}}{0.255},\ h_{\mathrm{pl}} \simeq \frac{M_{\mathrm{pl}}}{0.23},\ 0.392\lesssim\lambda_{\mathrm{DM}}(M_t)\lesssim0.395,\ y_\chi(M_t)\simeq0.5,\ \kappa(M_t)\simeq0.008$. The DM mass is $
901\ \mathrm{GeV}\lesssim M_{\chi} \lesssim972\  \mathrm{GeV}$.

However, $\lambda$ does not directly reach zero on the Planck scale, but it becomes negative just before the Planck scale. This does not solve the vacuum stability problem and it is the problem to be considered in the future.

Also, in the extended model of the SM that we considered in this study, we found that the DM becomes a WIMP-like DM by applying the MPP. These parameters successfully circumvent the current restrictions, affirm the WIMP, and are in good agreement with the WIMP's mass range. WIMPs cover the mass range around the Electroweak scale, which is extremely wide. Imposing the MPP conditions, it is possible to restrict the mass range to an extremely narrow range.

Finally, we briefly discuss the future prospects of our research.
Our extension is a simple extension, and other extensions may also be able to realize the MPP.
In short, if our model is excluded by experiments, it does not mean that the MPP has been eliminated in BSM.
For example, we can naively introduce gauge symmetry into the dark sector.
In addition, although the scalar field $S$ in our model has $Z_2$ symmetry, it would be so interesting to realize the MPP in a model that removes such symmetry.
Realizing the MPP in various models is so important in searches for the DM.
Additionally, it will be important to understand the theoretical meaning of the MPP.
%%%%%%%%%%%%%%%%%%%%%%%
%%%%%%%%%%%%%%%%%%%%%%%

\section*{Acknowledgments}
We would like to thank Noriaki Aibara, So Katagiri, Akio Sugamoto, and Ken Yokoyama for helpful comments. Especially, we are indebted to So Katagiri, Shiro Komata, and Akio Sugamoto for reading this paper and giving useful comments.

%%%%%%%%%%%%%%%%%%%%%%%
%%%%%%%%%%%%%%%%%%%%%%%
\newpage
\appendix
\def\thesection{Appendix \Alph{section}}
\section{One-loop Renormalization Group Equations}\label{firstappendix}
The one-loop RGEs in our Lagrangian (\ref{model}) are as follows:
\begin{eqnarray}
\frac{dg_Y}{dt}&=&\frac{1}{16\pi^2}\frac{41}{6}g_Y^3,\\ \frac{dg_2}{dt}&=&\frac{1}{16\pi^2}\frac{-19}{6}g_2^3,\\ \frac{dg_3}{dt}&=&\frac{-7}{16\pi^2}g_3^3, \\ 
\frac{dy_t}{dt}&=&\frac{y_t}{16\pi^2}\left(\frac{9}{2}y_t^2-\frac{17}{12}g_Y^2-\frac{9}4g_2^2-8g_3^2\right),\\
\frac{dy_\chi}{dt}&=&\frac{1}{16\pi^2}\frac{3}{2}y_\chi^3,\\
\frac{d\lambda}{dt} &=& \frac{1}{16\pi^2}\left(\lambda\left(24\lambda-3g_Y^2-9g_2^2+12y_t^2\right)+\frac{\kappa^2}{2}+\frac{3}{8}g_Y^4+\frac{3}{4}g_Y^2g_2^2+\frac{9}{8}g^4_2-6y_t^4\right),\\
\frac{d\lambda_{\mathrm{DM}}}{dt} &=& \frac{1}{16\pi^2}\left(\lambda_{\mathrm{DM}}\left(3\lambda_{\mathrm{DM}}+4y_\chi^2\right)+12\kappa^2-12y_\chi^4\right),\\
\frac{d\kappa}{dt} &=& \frac{1}{16\pi^2}\left(\kappa\left(-4\kappa+12\lambda+\lambda_{\mathrm{DM}}-\frac{3}2g_Y^2-\frac{9}2g_2^2+6y_t^2+2y_\chi^2\right)\right)
\end{eqnarray}
where $t$ is a certain parameter.
\def\thesection{Appendix B}
\section{Calculation of the annihilation cross section}
We calculate equation (\ref{CS_ex}) as an example of the formula for the annihilation cross section of the DM. In the case of non-relativistic DM, the annihilation cross section for $\chi\overline{\chi}\to f\overline{f}$ is given by
\begin{eqnarray}
(\sigma v_{\mathrm{rel}})_{\chi\overline{\chi}\to f\overline{f}}&=&\frac{1}{32\pi M_\chi^2}\left(1-\frac{M_f^2}{M_\chi^2}\right)^\frac{1}2\overline{|\mathcal{M}|^2}
\end{eqnarray}
with
\begin{eqnarray}
\overline{|\mathcal{M}|^2}_{\chi\overline{\chi}\to f\overline{f}}&=&\frac{1}{4}\times\frac{y_\chi^2y_f^2}{4}\left|\sum_{i=1,2}\theta_{f,i}\frac{\overline{v}_{\chi}(p_2)\theta_{\chi,i}u_\chi(p_1)}{\left(p_1+p_2\right)^2-M_{h_i}^2}\right|^2\left|\overline{u}_f(k_1)v_f(k_2)\right|^2\nonumber\\
&\simeq&\frac{y_\chi^2y_f^2}4(\sin2\theta)^2\left(2\left(M_\chi^2-M_f^2\right)+\frac{1}2M_\chi^2v_{\mathrm{rel}}^2\right)\nonumber\\
&\times&\left(2M_\chi^2+\frac{1}{2}M_\chi^2v_{\mathrm{rel}}^2\right)\times\left[\frac{1}{4M_\chi^2-M_{h_1}^2}-\frac{1}{4M_\chi^2-M_{h_2}^2}\right]^2
\end{eqnarray}
where $\theta_{\chi,1}=\theta_{f,2}=\cos\theta$ and $\theta_{\chi,2}=-\theta_{f,1}=\sin\theta$. The velocity of the DM $v_{\mathrm{rel}}$ is small.
%%%%%%%%%%%%%%%%%%%%%%%

%%%%%%%%%%%%%%%

%%%%%%%%%%%%%%%
%%%%%%%%%%%%
%%%%%%%%%%%%
\end{document}